\pdfoutput=1
\documentclass[a4paper,10pt,twocolumn,amsmath,amssymb,nofootinbib,%
superscriptaddress,groupedaddress,floats,floatfix,aps,prl]{revtex4-1}
\usepackage{graphicx,color}
\def\be#1\ee{\begin{align}#1\end{align}}
\newcommand{\ppar}{{p_\parallel}}

\newcommand{\lp}{\ell_{\rm p}}
\newcommand{\mpl}{m_{\rm p}}

\newcommand{\mb}{m_{\rm B}}
\begin{document}

\title{A new perspective on  galactic dynamics}

\date{\today}

\author{Mariano Cadoni}\email{mariano.cadoni@ca.infn.it}
\author{Matteo Tuveri}\email{matteo.tuveri@ca.infn.it}
\affiliation{Dipartimento di Fisica, Universit\`a di Cagliari
\& INFN, Sezione di Cagliari\\
Cittadella Universitaria, 09042 Monserrato, Italy}

\begin{abstract}
We derive the radial acceleration of stars in galaxies 
 by using basic features of thermodynamics, statistical 
mechanics and general relativity. 
We assume that the "dark" component of the radial 
acceleration is originated from the reaction of dark energy 
to the presence of baryonic matter. It can be also explained as 
the macroscopic manifestation of a huge number of 
extremely soft bosonic excitations of the dark energy medium 
with wavelength larger than the size of the cosmological horizon,
in thermal equilibrium with de Sitter spacetime.  
Our formula agrees with the phenomenological relation 
proposed by McGaugh {\sl et al.}  which, in turns, fits a 
large amount of observational data and with the MOND theory.
We also show that our formula appears as the 
weak field limit of  Einstein's 
general relativity sourced by an anisotropic fluid.
\end{abstract}

\maketitle%
The $\Lambda$CDM model of standard cosmology~\cite{Penzias:1965wn, Ade:2013zuv} 
accounts for present  experimental data 
about the  accelerated expansion of the universe~\cite{Riess:1998cb},
structure formation,  galaxy rotation curves and 
gravitational lensing
effects~\cite{Rubin:1980zd, Persic:1995ru, Massey:2010hh} by 
postulating the existence of  exotic forms of matter and 
energy (dark matter and dark energy (DE)).
Although the predictions of the $\Lambda$CDM 
model explain well the experimental data coming from large 
scale structure and cosmic microwave background, it fails to 
give a physical explanation of the so-called baryonic 
Tully-Fisher relation~\cite{Tully:1977fu,McGaugh:2000sr}, 
$v^2\simeq\sqrt{a_0\, G\, \mb}$. This formula relates the 
asymptotic velocity of stars in galaxies $v$ to the total 
baryonic mass $\mb$ through an acceleration parameter 
$a_0$ of the same order of magnitude 
of the current value of the Hubble constant $H$, $a_0\simeq 1.2\times 10^{-10} m\ s^{-2}$
~\cite{McGaugh:2016leg}.
Moreover, at galaxies and galaxy clusters level there are some 
mismatch between the $\Lambda$CDM predictions and observations 
~\cite{Klypin:1999uc,Moore:1999nt,BoylanKolchin:2011de,BoylanKolchin:2011dk}.

Motivated by these tensions, recently, there have been several attempts 
to explain the galactic-scale phenomenology commonly attributed to dark 
matter as due 
to a "dark force" (DF) generated by the reaction of DE to the 
presence of baryonic matter~\cite{Verlinde:2016toy,Cadoni:2017evg,Cadoni:2018dnd,Hossenfelder:2017eoh,Dai:2017guq,Cai:2017asf}. 
These approaches typically proceed by making a connection with 
phenomenological Milgrom's MOdified Newtonian Dynamics~(MOND)~\cite{Milgrom:1983ca,Milgrom:2014usa} 
in which $a_0$ is promoted to a fundamental constant of nature 
(see however~\cite{Rodrigues:2018duc,2018arXiv181205002C} 
for a critical discussions about this topic). 
The additional acceleration component of MOND, $a_{\rm MOND}=\sqrt{a_0a_{\rm B}}$, 
where $a_B$ is the Newtonian acceleration due to baryonic matter, 
is identified with that produced by the DF.  
In particular, from astrophysical observations we know  
that these DF effects arise when $a_B\simeq a_0$ 
to which corresponds a critical scale $r_0\simeq \sqrt{Gm_B/a_0}$.

The DF scenarios are powerful not only because
they avoid to postulate the existence of dark matter and give 
a physical explanation of the Tully-Fisher and MOND relation, 
but also because they explain why $a_0$ is a fundamental constant 
of the same order of magnitude of the cosmological acceleration $H$. 
On the other hand, they suffer from a serious drawback: 
when compared with observations, they are able to reproduce 
the asymptotic MOND formula but 
%differ significantly from observations 
 their predictions significantly differ from observations 
in the galactic region~
\cite{Milgrom:2016huh,Lelli:2017sul,Pardo:2017jun,Hees:2017uyk}.  
In order to fully reproduce the observational data about rotational 
curves of galaxies in the MOND framework, one needs to introduce a 
phenomenological interpolating function $F(x),\ x=a_B/a_0$, so that 
the total radial acceleration can be written as $a^r= F(x) a_B$~\cite{Milgrom:1983ca,McGaugh:2008nc,Sanders:2014xta}.
For $x\gg1$, near to the galactic core, the function $F$ must reproduce 
standard Newtonian gravity, $F(\infty)=1$, i.e. $a^r= a_B$,
where $a_B=Gm_B/r^2$. 
Instead, for $x\ll1$ we have the MOND regime, $F(x)\simeq \sqrt{\frac{1}{x}}$ 
and the radial acceleration is $a^r=a_{DF}=\sqrt{a_Ba_0} =\sqrt{a_0 Gm_B/r^2}$.
An interpolating function which satisfies all the conditions above 
and fits a large amount of observational data  coming from galaxies with 
different shapes (spiral, elliptical, spherical) has been proposed by 
McGaugh {\sl et al.}~\cite{McGaugh:2016leg,Lelli:2017vgz},~i.e. 
$F(x)= \frac{1}{1- e^{-\sqrt x}}$.

Observations of galactic dynamics imply that the total radial 
acceleration can be split in two components, the Newtonian 
contribution $a_B$ due to purely baryonic matter and an 
additional term to which we refer to as the DF 
contribution, $a_{DF}$, i.e. $a^r= a_B+a_{DF}$, with 
%{\mat What is evident from is that 
%the total radial acceleration , 
%the Newtonian contribution $a_B$ due to purely baryonic matter 
%and a "dark force" contribution $a_{DF}$. Therefore, the total 
%acceleration felt by stars in galaxies} 
%as $a^r= a_B+a_{DF}$, with 
%
\be\label{23}
a_{DF}= \frac{a_B}{e^{\sqrt\frac{a_B}{a_0}}-1}.
\ee
This formula shows that the total acceleration only depends on the 
baryonic matter distribution and, in principle, no dark matter is 
needed to fit the data~\cite{McGaugh:2016leg}.
In Eq.~\eqref{23}, $a_0$ appears as a fitting parameter and it 
is of the same order of magnitude of the cosmological acceleration.  
This allows us to write $a_0=\gamma H$, where $\gamma$ is a dimensionless 
parameter of order one. The value of $a_0$  found by McGaugh {\sl et al.} 
corresponds, approximatively, to $\gamma=1/2\pi$~\cite{McGaugh:2016leg}.
%Eq. \eqref{23} reproduces the Newtonian regime of gravity 
%for $a_B/a_0\to \infty$, giving $a_{DF}=0$ and $a^r= a_B$, 
%whereas the MOND regime, $a_{DF}=\sqrt{a_Ba_0} =\sqrt{a_0 Gm_B/r^2}$, 
%can be obtained as the limiting case $a_B/a_0\to 0$.
\medskip

The purpose of this letter is to show that in a DF scenario, 
Eq.~(\ref{23}) simply follows from basic features of thermodynamics, 
statistical mechanics and general relativity (GR) if one assumes 
that the DE medium responds in a simple and natural way to the 
presence of baryonic matter.  
We will also show that the same equation allows for a 
``metric-covariant uplifting'' as it appears as the weak-field limit of 
Einstein's general relativity sourced by an anisotropic fluid.
%

%Let us start by setting up our model. 
\section{The model}

We consider our universe as made only by DE and baryons.
We do not know what DE really is, 
thus we just consider the simpler case in which DE is described 
by a cosmological constant (see the discussion below for details). 
In absence of baryonic matter our universe is described, 
consistently with GR, by a de Sitter (dS) spacetime with a cosmological 
horizon $L$. 
%DE is simply given by a cosmological constant, 
%so that in absence of baryonic matter our universe is described, 
%consistently with GR, by a de Sitter (dS) spacetime with a cosmological 
%horizon $L$. 
The cosmological acceleration $H$ is related to $L$ by 
$H=1/L$\footnote{From now on we use natural units $c=\hbar=k_B=1$, 
whereas $\lp$, $\mpl=1/\lp$ and 
$G=\lp^2$ are the Planck length, the Planck mass and the Newton 
constant, respectively.}. 
The cosmological horizon, hence the dS spacetime, has an associated 
Bekenstein-Hawking temperature $T=1/(2\pi L)$~\cite{Narnhofer:1996zk,Deser:1997ri,Jacobson:1997ux}. 
We consider baryonic matter in the form of a point particle of mass $m_B$
\footnote{Our results can be easily generalized to the case of a spherically symmetric 
mass distribution $m_B(r)$.}
and its gravitational interaction with a 
test particle at distance $r$ from it in the weak field approximation. 
In a spherical region of radius $r$, this gravitational interaction will 
be given by the sum of the usual Newtonian component originated by the 
baryonic matter $m_B$ and of a DF component originated by the 
response of the DE to the presence of baryonic matter inside the sphere. 
DF effects arise when $r$ becomes comparable with the critical 
length scale  $r\simeq r_0$.
Hence, the total radial acceleration experienced by a test particle is 
$a^r=a_B+a_{DF}$.

%We assume the DF to be the macroscopic manifestation of bosonic 
%excitations of the DE medium (henceforth called ``DF bosonic excitations'') 
%with typical energy $\varepsilon\sim 1/r$.  
%We assume the DF to be the macroscopic manifestation of bosonic 
%excitations of the DE medium (henceforth called “DF bosonic excitations”) 
%with typical energy $\varepsilon\sim 1/r$. For instance, in a corpuscular 
%gravity scenario like that used in Ref.~\cite{Cadoni:2018dnd}, 
%these bosonic excitations can be considered as spin-2 particles (dark gravitons).
 A possible quantum description of the dS universe is that of  
a Bose-Einstein condensate of some quantum bosonic gravitational microscopic 
degrees of freedom~\cite{Cadoni:2019}.
In this context the DF has to be considered as the macroscopic manifestation of 
bosonic excitations of the DE medium (henceforth called "DF bosonic excitations") 
with typical energy $\varepsilon\sim 1/r$.
For instance, this is the case of a corpuscular gravity 
scenario like that used in Ref.~\cite{Cadoni:2018dnd} where these bosonic 
excitations can be considered as spin-2 particles (dark gravitons). 
In a thermodynamical, quantum mechanical picture, the DF acceleration 
$a_{DF}$ can be thought as generated by the
pressure $P$ of the gas of DF bosonic excitations in the sphere of 
radius $r$. Thus, we can write the acceleration for unit mass as 
$a_{DF}\sim P V/r \sim  PV \varepsilon$,  where $V$ is the volume 
of the sphere. 
At galactic scales, $r\ll L$, the 
thermal contribution, $TS$,  is negligible so that usual extensive 
thermodynamics implies $PV\sim  N\varepsilon$, where $N$ is the number of  
DF bosonic excitations in the sphere. This leads to
\be\label{11}
a_{DF}(\varepsilon)= C \varepsilon^2  N,
\ee   
where $C$ is a constant with  dimensions of a length,  whose 
value will be determined shortly. 
The additional DF is {\sl attractive} and has therefore the same 
sign of the Newtonian contribution. For simplicity, we  only consider the 
absolute value of forces and accelerations.
Notice that the scaling $a\sim \varepsilon^2 N$ given by Eq.~(\ref{11}) is 
the extensive counterpart of the sub-extensive  behaviour 
$a\sim  \varepsilon^2 \sqrt N$ found for the Newtonian 
term  $a_B$ in the radial acceleration~\cite{Mueck:2013mha,Cadoni:2017evg,Cadoni:2018dnd}. 
The extensive behaviour $V\sim  N,a\sim  N$ of our 
gas of DF bosonic excitations is perfectly consistent with 
its origin from the constant energy density characterizing 
the DE.

Assuming that the DF bosonic excitations are in thermal 
equilibrium with the dS spacetime and that their energy 
spectrum is non-degenerate, their 
number $N$ will follow a thermal Bose-Einstein distribution 
at a temperature $T=1/(2\pi L)$ and with zero chemical potential: 
$N(\varepsilon)= \frac{1}{e^{2\pi L\varepsilon}-1}$.

%{\mar The MOND regime appears when we have 
%a number $N >> 1$  of  soft DF bosonic excitations.  
%At leading order  in the  $2\pi L\varepsilon\to 0$ expansion 
%%we have  $N(\varepsilon)= (2\pi L \varepsilon)^{-1}$.
%Moreover,  this limit corresponds to $r\to L$ in which the universe 
%%is dominate by DE and the contribution of baryonic matter 
%can be neglected.  
%Thus we have $\varepsilon =\varepsilon_{DE}= 1/L$ and $a_{DF}$ must 
%become  the cosmological acceleration $a_{DF}= H$.}
The MOND regime appears at scales where the number of soft 
DF bosonic excitations becomes large, $N >> 1$, i.e. when 
$r_0\lesssim r<L$. In this limit, $2\pi L\varepsilon\to 0$ and, 
at leading order in $2\pi L\varepsilon$, we have 
$N(\varepsilon)= (2\pi L \varepsilon)^{-1}$. 
In the same limit the universe becomes DE dominated, being the 
contribution of baryonic matter negligible. 
The behaviour  of DF bosonic excitations should match 
that of the dS universe, i.e $\varepsilon =\varepsilon_{DE}= 1/L$ 
and $a_{DF}$ must become the cosmological acceleration 
$a_{DF}= H=1/L$~\cite{Cadoni:2017evg,Cadoni:2018dnd}.
This requirement determines the constant $C$ in Eq.~(\ref{11}) 
to be $C=2\pi L$. Putting all together we find
\be\label{19a}
a_{DF}(\varepsilon)= \frac{2\pi L \varepsilon^2} {e^{ 2\pi L\varepsilon}-1}.
\ee
%

%Our last task is the determination of the spectrum of soft DF gravitons, 
%i.e. of $\varepsilon$. 

\section{Determination of $\varepsilon$}

DF bosonic excitations are soft excitation of 
the DE medium.
The effect of introducing the baryonic matter will be the generation 
of a new effective interaction term between the baryonic mass $m_B $ 
and a test mass at distance $r$, which we have called DF. 
The energy $\varepsilon$ of the bosonic excitations 
mediating this dark interaction must 
therefore deviate from the simple Compton form $\varepsilon \sim 1/r$.
Introducing a dimensionless coupling constant $\alpha$ characterizing 
this interaction we write $\varepsilon \sim  \alpha /r$, where $\alpha$ 
can depend only on the quantities entering in the process, i.e. on $m_B,L, G$.

A direct determination of $\alpha$ requires the knowledge of 
the energy spectrum of the DF bosonic excitations. 
This in turns requires a detailed comprehension of the 
microphysics involved in the interaction between baryonic matter 
and DE.  Unfortunately, this is out of reach 
because we do not have any clear understanding of the microscopic 
nature of DE. 
 However, we can circumvent this problem by using, again, 
the macroscopic effect given in Eq.~(\ref{11}), i.e. 
the acceleration, produced by DF  excitation.  
Although the thermodynamical origin of this equation would 
require $N\gg1$, we can extrapolate its validity also for small 
values of $N$. In particular for $N=1$, Eq.~(\ref{11}) gives the DF 
acceleration $a_{DF}$ produced by a {\sl single} DF excitation. 
In view of the fact that $a_{DF}$ is the reaction of DE to baryonic  
matter, the simplest and most natural, assumption is that the DF 
acceleration produced by a single excitation is given exactly by 
what has generated it, i.e.~$a_{DF}=a_B$.  
With this assumption Eq.~(\ref{11}) gives
\be\label{oi}
\varepsilon=\sqrt\frac{G m_B}{2\pi L}\frac{1}{r}.
\ee
Using this result for $\varepsilon$, Eq.~(\ref{19a}) matches 
 with the phenomenological acceleration in Eq.~(\ref{23}), 
where the value of the dimensionless parameter, $\gamma=1/2\pi$, is  
perfectly compatible with the phenomenological value found 
by fitting a large amount of astrophysical data~\cite{McGaugh:2016leg,Lelli:2017vgz}.

Our model also predicts the correct value of the scale 
$r_0$ at which the dark force effects %set in
arise. 
This occurs when   $N(\varepsilon)\simeq 1$.
From the  Bose-Einstein distribution for 
$N(\varepsilon)$ 
we find that the condition above is satisfied when 
DF  modes  have energy $\varepsilon\simeq 1/L$,  which using  Eq.~(\ref{oi}) gives   
$r\simeq r_0=\sqrt{Gm_B/a_0}$.  
In the Newtonian regime, i.e. for $r\ll r_0$, we have hard DF 
excitations with energy $\varepsilon\gg 1/L$ whose 
number is exponentially suppressed and the DF effects are 
switched off. 
Finally, for $ r_0\lesssim r<L$, corresponding to the MOND regime, 
we have a huge number of extremely soft DF excitations, $N\gg 1$,
with energy $\varepsilon\ll 1/L$ and the DF effects are 
dominant. 

\section{Effective fluid description}

The relation in Eq.~(\ref{23}) can be also obtained as the weak field 
limit of a metric theory of gravity, namely 
GR sourced by an anisotropic fluid. This can be done along the lines 
of Ref.~\cite{Cadoni:2017evg}, where an effective fluid description 
is used and the radial pressure of the fluid describes the radial 
acceleration produced by the DF, $a^r=a_B+ 4\pi Gr \ppar$.
The pressure profile has to be chosen to match with Eq.~(\ref{23}):
\be\label{4}
\ppar(r)= \frac{1}{4\pi}\frac{m_B}{r^3}\frac{1}{e^{\frac{1}{r}\sqrt{Gm_B/a_0} }-1}.
\ee 
The full metric solution can be obtained solving Einstein field 
equations sourced by an anisotropic fluid with a pressure 
profile given by Eq.~(\ref{4}).  
The spacetime metric is taken of the form 
$ds^2= -f(r)e^{\Gamma(r)}dt^2+\frac{dr^2}{f(r)}+r^2d\Omega^2$ and 
it is given by
\be\label{6}
 f=1-\frac{2Gm_B}{r},\quad   
\Gamma'= \frac{2G}{r f}\left(
\frac{m_B}{r}\frac{1}{e^{\frac{1}{r}\sqrt{Gm_B/a_0} }-1}\right).
\ee
In the weak field limit of the metric one finds the potential 
$\phi= \frac{1}{2}(fe^\Gamma)$, from which one can easily derive 
the form of the DF 
%components for the 
acceleration and check 
that it exactly reproduces Eq.~(\ref{23}).
In the MOND regime of Eq.~(\ref{23}), i.e. $a_B/a_0\to 0$, 
one obtains the same expansion of the gravitational potential 
generated by a point-like particle given in Ref.~\cite{Cadoni:2017evg} 
with the characteristic logarithmic behaviour  of MOND
and an extremely 
tiny Machian contribution to the Newtonian 
potential~(see Ref.~\cite{Cadoni:2017evg}).

\section{Conclusions}

Starting from a DF scenario, using a simple and natural assumption 
for the reaction of DE to the presence of baryonic matter and basic 
features of thermodynamics and statistical mechanics, 
we have derived the phenomenological acceleration profile of stars in 
galaxies proposed by McGaugh {\sl et al.} in~\cite{McGaugh:2016leg,Lelli:2017vgz}.
Our formula reproduces both the Newtonian and MOND regimes 
of gravity as observed from the rotational curves data of a large 
variety of galaxies in the universe. 
Moreover, it can be also embedded in  GR.

The bosonic excitations %mediating 
responsible for the DF effects
are soft modes with wavelength of the order of the 
cosmological horizon, whose number  becomes 
significant only at galactic scales.
For this reason, they only affect the gravitational interaction 
at galactic scales, leaving unaltered the usual Newtonian 
contribution at smaller scales.
Our model also 
predict a value of the dimensionless acceleration parameter, 
$\gamma=1/(2\pi)$, which is in accordance with the phenomenological 
results in~\cite{McGaugh:2016leg,Lelli:2017vgz} obtained by 
fitting a large amount of observational data.
The same value has been obtained in alternative derivations linking $\gamma$ to 
the temperature of the dS spacetime ~\cite{Smolin:2017kkb,Alexander:2018lno}. 
Conversely, emergent gravity scenarios, which use area/volume competition effects, 
predict a slightly different value, $\gamma=1/6$~\cite{Verlinde:2016toy}. 
%%%%%%%%%%%%%%%%%%%%%%%%%%%%%%%%%%%%%%%%%%%%%%%%%%%%%%%%%%%%%%%%%%%%%%%%%%%%%%%%5

\section*{Aknowledgments}
We thank Roberto Casadio, Andrea Giusti and Wolfgang M\"{u}ck 
for valuable discussions and comments on this manuscript. 

%\bibliographystyle{utphys}
%\bibliography{gravitons}
%\bibliography{gravitonsDF}

\begin{thebibliography}{10}


\bibitem{Penzias:1965wn}
A.~A. Penzias and R.~W. Wilson, ``{A Measurement of excess antenna temperature
  at 4080-Mc/s},''
\href{http://dx.doi.org/10.1086/148307}{{\em Astrophys. J.} {\bfseries 142}
  (1965) 419--421}.
%%CITATION = ASJOA,142,419;%%.

\bibitem{Ade:2013zuv}
{\bfseries Planck} Collaboration, P.~A.~R. Ade {\em et~al.}, ``{Planck 2013
  results. XVI. Cosmological parameters},''
  \href{http://dx.doi.org/10.1051/0004-6361/201321591}{{\em Astron. Astrophys.}
  {\bfseries 571} (2014) A16},
\href{http://arxiv.org/abs/1303.5076}{{\ttfamily arXiv:1303.5076
  [astro-ph.CO]}}.
%%CITATION = ARXIV:1303.5076;%%.

\bibitem{Riess:1998cb}
{\bfseries Supernova Search Team} Collaboration, A.~G. Riess {\em et~al.},
  ``{Observational evidence from supernovae for an accelerating universe and a
  cosmological constant},'' \href{http://dx.doi.org/10.1086/300499}{{\em
  Astron. J.} {\bfseries 116} (1998) 1009--1038},
\href{http://arxiv.org/abs/astro-ph/9805201}{{\ttfamily arXiv:astro-ph/9805201
  [astro-ph]}}.
%%CITATION = ASTRO-PH/9805201;%%.

\bibitem{Rubin:1980zd}
V.~C. Rubin, N.~Thonnard, and W.~K. Ford, Jr., ``{Rotational properties of 21
  SC galaxies with a large range of luminosities and radii, from NGC 4605 /R =
  4kpc/ to UGC 2885 /R = 122 kpc/},''
\href{http://dx.doi.org/10.1086/158003}{{\em Astrophys. J.} {\bfseries 238}
  (1980) 471}.
%%CITATION = ASJOA,238,471;%%.

\bibitem{Persic:1995ru}
M.~Persic, P.~Salucci, and F.~Stel, ``{The Universal rotation curve of spiral
  galaxies: 1. The Dark matter connection},''
  \href{http://dx.doi.org/10.1093/mnras/281.1.27, 10.1093/mnras/278.1.27}{{\em
  Mon. Not. Roy. Astron. Soc.} {\bfseries 281} (1996) 27},
\href{http://arxiv.org/abs/astro-ph/9506004}{{\ttfamily arXiv:astro-ph/9506004
  [astro-ph]}}.
%%CITATION = ASTRO-PH/9506004;%%.

\bibitem{Massey:2010hh}
R.~Massey, T.~Kitching, and J.~Richard, ``{The dark matter of gravitational
  lensing},'' \href{http://dx.doi.org/10.1088/0034-4885/73/8/086901}{{\em Rept.
  Prog. Phys.} {\bfseries 73} (2010) 086901},
\href{http://arxiv.org/abs/1001.1739}{{\ttfamily arXiv:1001.1739
  [astro-ph.CO]}}.
%%CITATION = ARXIV:1001.1739;%%.

\bibitem{Tully:1977fu}
R.~B. Tully and J.~R. Fisher, ``{A New method of determining distances to
  galaxies},''
{\em Astron. Astrophys.} {\bfseries 54} (1977) 661--673.
%%CITATION = AAEJA,54,661;%%.

\bibitem{McGaugh:2000sr}
S.~S. McGaugh, J.~M. Schombert, G.~D. Bothun, and W.~J.~G. de~Blok, ``{The
  Baryonic Tully-Fisher relation},''
  \href{http://dx.doi.org/10.1086/312628}{{\em Astrophys. J.} {\bfseries 533}
  (2000) L99--L102},
\href{http://arxiv.org/abs/astro-ph/0003001}{{\ttfamily arXiv:astro-ph/0003001
  [astro-ph]}}.
%%CITATION = ASTRO-PH/0003001;%%.

\bibitem{McGaugh:2016leg}
S.~McGaugh, F.~Lelli, and J.~Schombert, ``{Radial Acceleration Relation in
  Rotationally Supported Galaxies},''
  \href{http://dx.doi.org/10.1103/PhysRevLett.117.201101}{{\em Phys. Rev.
  Lett.} {\bfseries 117} no.~20, (2016) 201101},
\href{http://arxiv.org/abs/1609.05917}{{\ttfamily arXiv:1609.05917
  [astro-ph.GA]}}.
%%CITATION = ARXIV:1609.05917;%%.

\bibitem{Klypin:1999uc}
A.~A. Klypin, A.~V. Kravtsov, O.~Valenzuela, and F.~Prada, ``{Where are the
  missing Galactic satellites?},'' \href{http://dx.doi.org/10.1086/307643}{{\em
  Astrophys. J.} {\bfseries 522} (1999) 82--92},
\href{http://arxiv.org/abs/astro-ph/9901240}{{\ttfamily arXiv:astro-ph/9901240
  [astro-ph]}}.
%%CITATION = ASTRO-PH/9901240;%%.

\bibitem{Moore:1999nt}
B.~Moore, S.~Ghigna, F.~Governato, G.~Lake, T.~R. Quinn, J.~Stadel, and
  P.~Tozzi, ``{Dark matter substructure within galactic halos},''
  \href{http://dx.doi.org/10.1086/312287}{{\em Astrophys. J.} {\bfseries 524}
  (1999) L19--L22},
\href{http://arxiv.org/abs/astro-ph/9907411}{{\ttfamily arXiv:astro-ph/9907411
  [astro-ph]}}.
%%CITATION = ASTRO-PH/9907411;%%.

\bibitem{BoylanKolchin:2011de}
M.~Boylan-Kolchin, J.~S. Bullock, and M.~Kaplinghat, ``{Too big to fail? The
  puzzling darkness of massive Milky Way subhaloes},''
  \href{http://dx.doi.org/10.1111/j.1745-3933.2011.01074.x}{{\em Mon. Not. Roy.
  Astron. Soc.} {\bfseries 415} (2011) L40},
\href{http://arxiv.org/abs/1103.0007}{{\ttfamily arXiv:1103.0007
  [astro-ph.CO]}}.
%%CITATION = ARXIV:1103.0007;%%.

\bibitem{BoylanKolchin:2011dk}
M.~Boylan-Kolchin, J.~S. Bullock, and M.~Kaplinghat, ``{The Milky Way's bright
  satellites as an apparent failure of LCDM},''
  \href{http://dx.doi.org/10.1111/j.1365-2966.2012.20695.x}{{\em Mon. Not. Roy.
  Astron. Soc.} {\bfseries 422} (2012) 1203--1218},
\href{http://arxiv.org/abs/1111.2048}{{\ttfamily arXiv:1111.2048
  [astro-ph.CO]}}.
%%CITATION = ARXIV:1111.2048;%%.

\bibitem{Verlinde:2016toy}
E.~P. Verlinde, ``{Emergent Gravity and the Dark Universe},''
  \href{http://dx.doi.org/10.21468/SciPostPhys.2.3.016}{{\em SciPost Phys.}
  {\bfseries 2} no.~3, (2017) 016},
\href{http://arxiv.org/abs/1611.02269}{{\ttfamily arXiv:1611.02269 [hep-th]}}.
%%CITATION = ARXIV:1611.02269;%%.

\bibitem{Cadoni:2017evg}
M.~Cadoni, R.~Casadio, A.~Giusti, W.~Mueck, and M.~Tuveri, ``{Effective Fluid
  Description of the Dark Universe},''
  \href{http://dx.doi.org/10.1016/j.physletb.2017.11.058}{{\em Phys. Lett.}
  {\bfseries B776} (2018) 242--248},
\href{http://arxiv.org/abs/1707.09945}{{\ttfamily arXiv:1707.09945 [gr-qc]}}.
%%CITATION = ARXIV:1707.09945;%%.

\bibitem{Cadoni:2018dnd}
M.~Cadoni, R.~Casadio, A.~Giusti, and M.~Tuveri, ``{Emergence of a Dark Force
  in Corpuscular Gravity},''
  \href{http://dx.doi.org/10.1103/PhysRevD.97.044047}{{\em Phys. Rev.}
  {\bfseries D97} no.~4, (2018) 044047},
\href{http://arxiv.org/abs/1801.10374}{{\ttfamily arXiv:1801.10374 [gr-qc]}}.
%%CITATION = ARXIV:1801.10374;%%.

\bibitem{Hossenfelder:2017eoh}
S.~Hossenfelder, ``{Covariant version of Verlinde's emergent gravity},''
  \href{http://dx.doi.org/10.1103/PhysRevD.95.124018}{{\em Phys. Rev.}
  {\bfseries D95} no.~12, (2017) 124018},
\href{http://arxiv.org/abs/1703.01415}{{\ttfamily arXiv:1703.01415 [gr-qc]}}.
%%CITATION = ARXIV:1703.01415;%%.

\bibitem{Dai:2017guq}
D.-C. Dai and D.~Stojkovic, ``{Comment on 'Covariant version of Verlinde's
  emergent gravity'},''
  \href{http://dx.doi.org/10.1103/PhysRevD.96.108501}{{\em Phys. Rev.}
  {\bfseries D96} no.~10, (2017) 108501},
\href{http://arxiv.org/abs/1706.07854}{{\ttfamily arXiv:1706.07854 [gr-qc]}}.
%%CITATION = ARXIV:1706.07854;%%.

\bibitem{Cai:2017asf}
R.-G. Cai, S.~Sun, and Y.-L. Zhang, ``{Emergent Dark Matter in Late Universe on
  Holographic Screen},''
\href{http://arxiv.org/abs/1712.09326}{{\ttfamily arXiv:1712.09326 [hep-th]}}.
%%CITATION = ARXIV:1712.09326;%%.

\bibitem{Milgrom:1983ca}
M.~Milgrom, ``{A Modification of the Newtonian dynamics as a possible
  alternative to the hidden mass hypothesis},''
\href{http://dx.doi.org/10.1086/161130}{{\em Astrophys. J.} {\bfseries 270}
  (1983) 365--370}.
%%CITATION = ASJOA,270,365;%%.

\bibitem{Milgrom:2014usa}
M.~Milgrom, ``{MOND theory},''
  \href{http://dx.doi.org/10.1139/cjp-2014-0211}{{\em Can. J. Phys.} {\bfseries
  93} no.~2, (2015) 107--118},
\href{http://arxiv.org/abs/1404.7661}{{\ttfamily arXiv:1404.7661
  [astro-ph.CO]}}.
%%CITATION = ARXIV:1404.7661;%%.

\bibitem{Rodrigues:2018duc}
D.~C. Rodrigues, V.~Marra, A.~del Popolo, and Z.~Davari, ``{Absence of a
  fundamental acceleration scale in galaxies},''
  \href{http://dx.doi.org/10.1038/s41550-018-0498-9}{{\em Nat. Astron.}
  {\bfseries 2} no.~8, (2018) 668--672},
\href{http://arxiv.org/abs/1806.06803}{{\ttfamily arXiv:1806.06803
  [astro-ph.GA]}}.
%%CITATION = ARXIV:1806.06803;%%.

\bibitem{2018arXiv181205002C}
Z.~{Chang} and Y.~{Zhou}, ``{Is there a fundamental acceleration scale in
  galaxies?},'' {\em arXiv e-prints} (Dec., 2018) arXiv:1812.05002,
  \href{http://arxiv.org/abs/1812.05002}{{\ttfamily arXiv:1812.05002
  [astro-ph.GA]}}.

\bibitem{Milgrom:2016huh}
M.~Milgrom and R.~H. Sanders, ``{Perspective on MOND emergence from Verlinde's
  "emergent gravity" and its recent test by weak lensing},''
\href{http://arxiv.org/abs/1612.09582}{{\ttfamily arXiv:1612.09582
  [astro-ph.GA]}}.
%%CITATION = ARXIV:1612.09582;%%.

\bibitem{Lelli:2017sul}
F.~Lelli, S.~S. McGaugh, and J.~M. Schombert, ``{Testing Verlinde's Emergent
  Gravity with the Radial Acceleration Relation},''
  \href{http://dx.doi.org/10.1093/mnrasl/slx031}{{\em Mon. Not. Roy. Astron.
  Soc.} {\bfseries 468} no.~1, (2017) L68--L71},
\href{http://arxiv.org/abs/1702.04355}{{\ttfamily arXiv:1702.04355
  [astro-ph.GA]}}.
%%CITATION = ARXIV:1702.04355;%%.

\bibitem{Pardo:2017jun}
K.~Pardo, ``{Testing Emergent Gravity with Isolated Dwarf Galaxies},''
\href{http://arxiv.org/abs/1706.00785}{{\ttfamily arXiv:1706.00785
  [astro-ph.CO]}}.
%%CITATION = ARXIV:1706.00785;%%.

\bibitem{Hees:2017uyk}
A.~Hees, B.~Famaey, and G.~Bertone, ``{Emergent gravity in galaxies and in the
  Solar System},'' \href{http://dx.doi.org/10.1103/PhysRevD.95.064019}{{\em
  Phys. Rev.} {\bfseries D95} no.~6, (2017) 064019},
\href{http://arxiv.org/abs/1702.04358}{{\ttfamily arXiv:1702.04358
  [astro-ph.GA]}}.
%%CITATION = ARXIV:1702.04358;%%.

\bibitem{McGaugh:2008nc}
S.~McGaugh, ``{Milky Way Mass Models and MOND},''
  \href{http://dx.doi.org/10.1086/589148}{{\em Astrophys. J.} {\bfseries 683}
  (2008) 137--148},
\href{http://arxiv.org/abs/0804.1314}{{\ttfamily arXiv:0804.1314 [astro-ph]}}.
%%CITATION = ARXIV:0804.1314;%%.

\bibitem{Sanders:2014xta}
R.~H. Sanders, ``{A historical perspective on modified Newtonian dynamics},''
  \href{http://dx.doi.org/10.1139/cjp-2014-0206}{{\em Can. J. Phys.} {\bfseries
  93} no.~2, (2015) 126--138},
\href{http://arxiv.org/abs/1404.0531}{{\ttfamily arXiv:1404.0531
  [physics.hist-ph]}}.
%%CITATION = ARXIV:1404.0531;%%.

\bibitem{Lelli:2017vgz}
F.~Lelli, S.~S. McGaugh, J.~M. Schombert, and M.~S. Pawlowski, ``{One Law to
  Rule Them All: The Radial Acceleration Relation of Galaxies},''
  \href{http://dx.doi.org/10.3847/1538-4357/836/2/152}{{\em Astrophys. J.}
  {\bfseries 836} no.~2, (2017) 152},
\href{http://arxiv.org/abs/1610.08981}{{\ttfamily arXiv:1610.08981
  [astro-ph.GA]}}.
%%CITATION = ARXIV:1610.08981;%%.

\bibitem{Narnhofer:1996zk}
H.~Narnhofer, I.~Peter, and W.~E. Thirring, ``{How hot is the de Sitter
  space?},'' \href{http://dx.doi.org/10.1142/S0217979296000611}{{\em Int. J.
  Mod. Phys.} {\bfseries B10} (1996) 1507--1520}.
[,603(1996)].
%%CITATION = IMPAE,B10,1507;%%.

\bibitem{Deser:1997ri}
S.~Deser and O.~Levin, ``{Accelerated detectors and temperature in (anti)-de
  Sitter spaces},'' \href{http://dx.doi.org/10.1088/0264-9381/14/9/003}{{\em
  Class. Quant. Grav.} {\bfseries 14} (1997) L163--L168},
\href{http://arxiv.org/abs/gr-qc/9706018}{{\ttfamily arXiv:gr-qc/9706018
  [gr-qc]}}.
%%CITATION = GR-QC/9706018;%%.

\bibitem{Jacobson:1997ux}
T.~Jacobson, ``{Comment on `Accelerated detectors and temperature in anti-de
  Sitter spaces'},'' \href{http://dx.doi.org/10.1088/0264-9381/15/1/020}{{\em
  Class. Quant. Grav.} {\bfseries 15} (1998) 251--253},
\href{http://arxiv.org/abs/gr-qc/9709048}{{\ttfamily arXiv:gr-qc/9709048
  [gr-qc]}}.
%%CITATION = GR-QC/9709048;%%.

\bibitem{Mueck:2013mha}
W.~Mueck, ``{On the number of soft quanta in classical field configurations},''
  \href{http://dx.doi.org/10.1139/cjp-2013-0712}{{\em Can. J. Phys.} {\bfseries
  92} no.~9, (2014) 973--975},
\href{http://arxiv.org/abs/1306.6245}{{\ttfamily arXiv:1306.6245 [hep-th]}}.
%%CITATION = ARXIV:1306.6245;%%.

\bibitem{Cadoni:2019}
M.~Cadoni and M.~Tuveri, "{In preparation.}"


\bibitem{Smolin:2017kkb}
L.~Smolin, ``{MOND as a regime of quantum gravity},''
  \href{http://dx.doi.org/10.1103/PhysRevD.96.083523}{{\em Phys. Rev.}
  {\bfseries D96} no.~8, (2017) 083523},
\href{http://arxiv.org/abs/1704.00780}{{\ttfamily arXiv:1704.00780 [gr-qc]}}.
%%CITATION = ARXIV:1704.00780;%%.

\bibitem{Alexander:2018lno}
S.~Alexander and L.~Smolin, ``{The Equivalence Principle and the Emergence of
  Flat Rotation Curves},''
\href{http://arxiv.org/abs/1804.09573}{{\ttfamily arXiv:1804.09573 [gr-qc]}}.
%%CITATION = ARXIV:1804.09573;%%.

\end{thebibliography}
%
%\providecommand{\href}[2]{#2}\begingroup\raggedright
%\endgroup

\end{document}